\documentclass[aps,amsmath,amssymb,preprintnumbers,groupedaddress,twocolumn]{revtex4}
\usepackage{graphicx,psfrag}

\newcommand{\M}{\mathcal M}

\renewcommand{\c}{\overline}

\newcommand{\beq}{\begin{equation}}
\newcommand{\eq}{\end{equation}}

\newcommand{\U}{\widetilde{U(1)}}
\renewcommand{\O}{\widetilde{O(1)}}

\newcommand{\SP}{\widetilde{Sp(2)}}

\begin{document}

\preprint{CERN-PH-TH/2008-063}\preprint{LMU-ASC 14/08}
\title{A gauge theory analog of some ``stringy" D-instantons}

\author{Daniel Krefl}
\affiliation{\\\\Arnold Sommerfeld Center for Theoretical Physics,\\Ludwig-Maximilians-Universit\"at, Theresienstr. 37, 80333 Munich, Germany \\\\ and \\\\$CERN,~PH$-$TH~Division,~1211~Geneva, ~Switzerland $}

\begin{abstract}
We argue that one can see a specific class of ``stringy" D-instantons in the underlying 4D gauge theory as the UV completion of an ordinary gauge instanton of a completely broken gauge group corresponding to the ``empty" cycle the D-instanton is located on. In this sense, the D-instanton induced non-perturbative superpotential can be qualitatively inferred directly from pure field-theory considerations.
\end{abstract}
\maketitle
\noindent

\section{Introduction}

During recent times it became clear that D-branes which are wrapped only on internal cycles which are not occupied by other D-branes living as well in space-time, are quite of interest for model building via Type II intersecting brane setups \cite{Blumenhagen:2006xt,Ibanez:2006da,Florea:2006si}. 
Such D-branes are usually refered to as ``stringy" instantons. Especially, they can lead to phenomenologically desirable couplings as for example Majorana mass terms \cite{Blumenhagen:2006xt,Ibanez:2006da} or Yukawa couplings \cite{Blumenhagen:2007zk} and turned out as well to be useful for constructing models which feature dynamically supersymmetry breaking \cite{Argurio:2007qk,Aganagic:2007py,Aharony:2007db,Buican:2007is}. 
The conditions under which a non-perturbative contribution to the low-energy effective gauge theory can be generated by a ``stringy" instanton are quite restrictive. In detail, besides two fermionic zero modes of the uncharged sector, all additional fermionic zero-modes must be saturated. Especially, without any extra ingredients (like multi-instanton effects) only contributions from $\O$ instantons \cite{Argurio:2007vqa,Ibanez:2007rs} and, as recently found, if the cycle is occupied in addition by a single D-brane living in space-time as well, from $\U$ instantons \cite{Petersson:2007sc} are possible. Here and in the following we denote by a $\widetilde{~~~~~~}$ the world-volume gauge theory of the instanton brane. 

However, up to now it is not clear in what sense one should interpret these ``stringy" instantons from a gauge theory point of view or if there is an interpretation at all. Especially, the observation that one can transform in specific setups a ``stringy" instanton to an ordinary gauge theory effect is puzzling \cite{Aharony:2007pr}, since it implies that there should be indeed an analog of ``stringy" instantons in gauge theory, in contrast to otherwise statements in the literature.

The purpose of this note is to shed some more light onto this puzzle. For that, we will first consider the simplest imaginable setup involving an $\U$ instanton, and as will turn out, these ``stringy" instantons can be as well dualized to gauge theory strong coupling effects. In fact, as we will argue, the underlying point is that one can see some specific ``stringy" instantons to be analog to gauge theory instantons of completely broken gauge groups \footnote{These instantons are also known as ``residual instantons".}. In more detail, via higgsing one can interpolate between both, there for small vevs the gauge theory picture of a completely broken gauge group is valid, while for vevs around the string scale the gauge theory picture breaks down, however stringy physics in form of the ``stringy" instanton jumps in and reproduces an analog non-perturbative potential and hence ensures continuity of the instanton effect. In this sense, the qualitative form of the non-perturbative effect is determined already by gauge theory considerations. Especially, the phenomenological implications of these non-perturbative effects should be analog to the ones considered already some time ago in pure field-theory (see for instance \cite{Csaki:1996gr}). 

Finally, we will shortly comment that the same considerations hold as well for $\O$ instantons and conclude by giving an example of a ``stringy" $\SP$ instanton.

\section{$\U\times U(1)$ theory}

Let us consider the recently proposed case of a single euclidean D-brane with $\U$ world-volume theory placed on a single space-time filling D-brane, which has been claimed to be a ``stringy" D-instanton effect \cite{Petersson:2007sc}.

\begin{figure}[!h]
\psfrag{A}[cc][][0.8]{$NS'$}
\psfrag{B}[cc][][0.8]{$NS$}
\psfrag{C}[cc][][0.8]{$N_f~D6$}
\psfrag{D}[cc][][0.8]{$D4$}
\psfrag{E}[cc][][0.8]{$N_f~D4$}
\psfrag{F}[cc][][0.8]{$E0$}
\begin{center}
\includegraphics[scale=0.3]{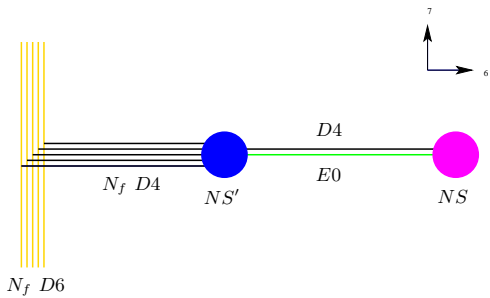}
\caption{HW setup involving a $\U$ ``stringy" instanton.}
\label{U1fig1}
\end{center}
\end{figure}

One of the simplest imaginable D-brane setups involving such a $\U$ ``stringy" D-instanton is the following classic Hanany-Witten (for short HW) setup:
We place in the $x^6$ direction from left to right, a stack of $N_f$ $D6$-branes which live along $(*|x^7,x^8,x^9)$, a $NS'$-brane living along $(*|x^8,x^9)$ and a $NS$ brane living along $(*|x^4,x^5)$, where $*$ stands for the space-time coordiantes $(x^0,x^1,x^2,x^3)$. The stack of $D6$-branes is connected by $N_f$ flavor $D4$-branes living in $(*|[x^6])$,  to the $NS'$-brane and the $NS'$-brane is connected by a single color $D4$-brane and an instanton $E0$-brane to the $NS$-brane. The $E0$ brane lives only in a finite interval of $x^6$, i.e. $(|[x^6])$. For convenience, the setup is illustrated in figure \ref{U1fig1}.

The reader familar with HW setups immediately identifies the corresponding low-energy theory to be a non-existing ``$SU(1)$" (in the sense that the single $U(1)$ brane corresponds to a global symmetry) with a meson $\M_{ij}$ and quarks $q_i$ and $\c q_i$ (here and in the following the indices are always flavor indices). Further, there will be a superpotential
\beq\label{U1eq1}
W_{cl}=q\M \c q,
\eq
where the summation over flavor indices is implicitly understood.
Besides the usual fermionic and bosonic zero modes of an $E0$-brane on top of a $D4$-brane which are taken care of by the ADHM constraints, as described in detail in \cite{Petersson:2007sc}, we have in addition $N_f$ fermionic zero-modes $\lambda_i$ and $\c\lambda_i$ from the intersection with the $N_f$ flavor $D4$-branes, which need to be taken care of. The superpotential (\ref{U1eq1}) immediately tells us that these zero-modes are lifted via the coupling $\lambda_i\M_{ij}\c\lambda_j$ \footnote{This is implicit from \cite{Franco:2007ii}.}. Thus, the effect of the $E0$-brane is to yield a non-perturbative superpotential of the form
\beq\label{U1eq2}
W_{np}\sim \det(\M).
\eq
The form of the combined superpotential $W=W_{cl}+W_{np}$ implies that we should see this theory as the magnetic dual of an electric $SU(N_c)$ theory with $N_f=N_c+1$ flavors, which undergoes s-confinement and where the combined superpotential arises due to strong coupling dynamics.  

Let us check if the dynamical scale of (\ref{U1eq2}) fits into this picture:
One can show that the dynamical scale of this non-perturbative potential is given by (cf. \cite{Petersson:2007sc,Billo:2002hm})
\beq\label{scale}
W_{np}\sim\tilde \Lambda^{3-N_f},
\eq 
where
$\tilde\Lambda^{3-N_f}:=M_s^{3-N_f}e^{\langle1\rangle^{disc}}$ with $M_s$ the string scale and $\langle1\rangle^{disc}$ the vacuum disc amplitude of the instanton. Using the well-known field-theory matching relation between the scales of the magnetic and electric theory $\tilde\Lambda^{3-N_f}\Lambda^{2N_c-1}=-\mu^{N_f}$, where $\mu$ is an additional scale (in our case naturally the string scale) relating the mesons of the magnetic and electric theory, i.e. $\M_{el}=\mu\M$, we see that in the dual electric theory the scaling of (\ref{U1eq2}) is as expected from field-theory:
\beq
W^{el}_{np}\sim-\frac{1}{\Lambda^{2N_c-1}}.
\eq

Indeed, exchaning the position of the $NS$- and $NS'$-brane and moving subsequently the stack of $D6$-branes between them, yields the expected dual electric theory. Note that in this process the $E0$-brane switches orientation and hence the instanton is converted to an anti-instanton (however there is no net change since at the same time an anti-instanton is converted to a instanton). 

It is clear that this behavior holds for any HW setup involving an $E0$-brane on top of a single $D4$-brane which yields a non-trivial superpotential, i.e. we can always interpret this in the way that we have in our theory a s-confining $SU(N_c)$ gauge factor with $N_f=N_c+1$.

Thus, also for the $\U$ instantons we have rediscovered the known fact that one can convert a ``stringy" instanton which yields a non-perturbative potential to some gauge theory strong coupling dynamics \cite{Aharony:2007pr,GarciaEtxebarria:2007zv}. This immediately leads us to a paradox, since a ``stringy" instanton is widely believed to be a pure string effect and then it should not be possible to transform it  to some gauge dynamics.

In order to resolve this puzzle, let us take a closer look on the field-theory side: In pure field-theory we cannot perform directly this duality, but rather need to consider an electric $SU(N_c)$ theory with one massive and $N_f=N_c+1$ massless flavors, which can be seen as an UV completion of the pure $N_f=N_c+1$ theory. While the mass-term in the electric theory reduces simply the flavor number by one and gives us the theory we are interested in, the effect of the mass-term in the dual theory is to break completely the dual $SU(2)$ group. The non-perturbative superpotential can then be seen to arise from the instanton of the completely broken $SU(2)$ group \cite{Seiberg:1994pq}. The scale of the breaking is given by the scale of the vev of the field which breaks the gauge group, which is in turn determined by the mass term. Especially, the scale of the resulting non-perturbative potential is inversely proportional to the scale of the vev. 

In contrast, the string theory brane setup is UV complete, so we can perform directly the field-theory duality in the brane picture. In this case however, the ``stringy" instanton mimics the gauge instanton of the completely broken gauge group we had in the pure field-theory discussion. In order to illustrate this, let us place an additional $D4$-brane onto the $E0$-brane in our original HW setup and as well an additional $D6$-brane onto the $D6$-stack and connect it with an additional flavor $D4$-brane to the $NS$-brane. Clearly, the $E0$-brane corresponds to a gauge instanton in the $SU(2)$ world-volume theory of the two $D4$-branes. However, we can easily remove again the additional introduced branes by reconnecting the additional color $D4$-brane with one of the flavor $D4$-branes and moving the $D6$-brane the reconnected $D4$-brane ends on of along the $(x^4,x^5)$ direction. In field-theory, this can be seen as breaking the $SU(2)$ group by a flavor vev induced by a linear superpotential term. For small vevs, this exactly mimics the original field-theory discussions. For vevs around the string scale the field-theory description breaks down and stringy physics comes into the game. However, for string scale vevs we can as well safely remove the branes to infinity, and obtain our original ``stringy" instanton setup, which yields the qualitatively same superpotential as the gauge theory picture for small vevs. As is now clear, we can continuously interpolate between a gauge theory instanton and a ``stringy" instanton via breaking the corresponding gauge group. In this sense, the ``stringy" instanton can be seen as an UV completition of the instanton of a completely broken gauge group. Let us see how the scale (\ref{U1eq2}) fits into this picture: Bringing in suitable powers of $M_s$, we can rewrite (\ref{U1eq2}) as
\beq
\tilde\Lambda^{3-N_f}=\frac{\tilde \Lambda^{6-(N_f+1)}}{M_s^2}=-\frac{\tilde \Lambda^{6-(N_f+1)}}{\langle q\bar q\rangle},
\eq
thus, we can not distinguish the ``stringy" instanton scale (lhs) from the scale of the instanton of a $SU(2)$ completely broken at the string scale by the vev $\langle q\bar q\rangle=-\mu m=-M_s^2$ (rhs), induced by a linear term in the superpotential proportional to $m$.
 
The upshot is, since it is well known that an $E0$ brane inside a $D4$ brane corresponds to a gauge instanton in the world-volume theory of the $D4$-brane and that if we remove the higher dimensional brane in what ever way (including the case that it has not been present at all), and keep just the instanton D-brane, then we can see the ``stringy" D-instanton as a left-over of the gauge instanton of the removed higher dimensional brane. In other words, in the low-energy field-theory we can interpret the ``empty" cycle as a completely broken gauge group (at the string scale) and the ``stringy" D-instanton as the gauge instanton thereof. Thus, the qualitative contribution of a ``stringy" D-instanton can then as well be inferred purely from field-theory considerations.

\begin{figure}[!h]
\psfrag{A}[cc][][0.8]{$E2+D6$}
\psfrag{B}[cc][][0.8]{$D6_L$}
\psfrag{C}[cc][][0.8]{$D6_R$}
\psfrag{D}[cc][][0.8]{$D6_1...D6_N$}
\psfrag{Y}[cc][][0.8]{$E2+D6+D6'$}
\psfrag{c}[cc][][0.8]{$\M_{ij}$}
\psfrag{a}[cc][][0.8]{$q_i,\lambda_i$}
\psfrag{b}[cc][][0.8]{$\c q_i,\c \lambda_i$}
\psfrag{z}[cc][][0.8]{higgsing}
\begin{center}
\includegraphics[scale=0.25]{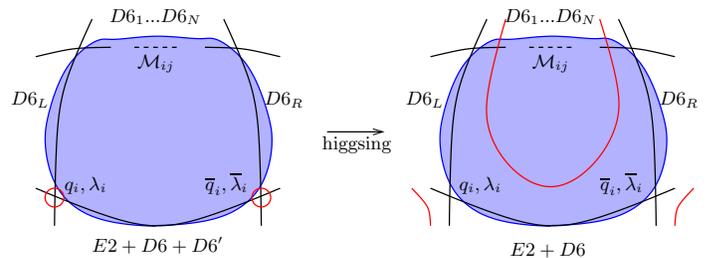}
\caption{Type IIA intersecting $D6$-brane setup involving a $\U$ ``stringy" instanton, obtained from higgsing from a brane setup involving an ordinary gauge theory instanton.}
\label{U1fig2}
\end{center}
\end{figure}

In order to illustrate that this picture does not only apply for local setups where one can move of the branes to infinity, let us switch gears and consider the simplest imaginable intersecting $D6$-brane setup involving an $\U$ instanton. In detail, we consider an $E2$-brane on top of two $D6$-branes, denoted as $D6$ and $D6'$, intersected by two other stacks of $D6$-branes $D6_L$ and $D6_R$ and a set of $N$ additional stacks of $D6$-branes intersecting in a way such that we obtain a holomorphic disk involving a $D6_L$-$E2$-$D6_R$ boundary component. We can now always give a vev to the quark fields located at the $D6_L-D6'$ and $D6'-D6_R$ intersection and higgs in this way this gauge theory instanton to a pure ``stringy" instanton. The additional $D6$ brane that has been present before the higgsing recombines with a pair of flavor branes to a brane which does not anymore intersect the setup and thus forms a hidden sector. This is illustrated in figure \ref{U1fig2}.

After the higgsing, the low-energy effective gauge theory of this setup is given by
\beq
... \times SU(N_L)\times ``SU(1)"\times SU(N_R)\times ..., 
\eq
where the dots stand for the gauge groups of the other D-branes which are not important for our discussion and we keep for illustration explicitly the left-over non-existing ``$SU(1)$" gauge factor of the single $D6$-brane (where one should see again the single $U(1)$ brane as providing a global symmetry). As in the HW example considered previously, the trivial factor has quarks $q_i,\c q_j$ and a meson $\M_{ij}$ and we have the superpotentials given in equations (\ref{U1eq1}) and (\ref{U1eq2}) arising via a world-sheet disk instanton and respectively a ``stringy" D-instanton.

For the same reasons as before, from the 4D effective field-theory point of view this trivial gauge factor with its ``stringy" instanton induced superpotential looks like the magnetic dual of a s-confining theory.  In an intersecting $D6$-brane setup, it is not so clear how one could perform explicitly this duality, however we expect that the duality arises similar as in the non-compact case considered in \cite{Hori:2000ck} via motion in complex structure moduli space. Anyway, the point is, that we can qualitatively reproduce the effect of the ``stringy" instanton in field-theory by seeing the trivial $``SU(1)"$ factor as arising from a completely broken $SU(2)$ group, as described via the higgsing procedure above by just giving a vev below the string scale. Thus, in the low-energy effective field-theory we cannot really distinguish between a non-perturbative superpotential coming from a ``stringy" D-instanton and a superpotential coming from the instanton of a completely broken $SU(2)$.

\section{$\O$ instantons}
\begin{figure}[!h]
\psfrag{A}[cc][][0.8]{$E2$}
\psfrag{B}[cc][][0.8]{$D6_L$}
\psfrag{C}[cc][][0.8]{$D6_R'$}
\psfrag{D}[cc][][0.8]{$D6_1...D6_N$}
\psfrag{c}[cc][][0.8]{$\M_{ij}$}
\psfrag{a}[cc][][0.8]{$\lambda_i$}
\psfrag{b}[cc][][0.8]{$\c \lambda_i'$}
\begin{center}
\includegraphics[scale=0.25]{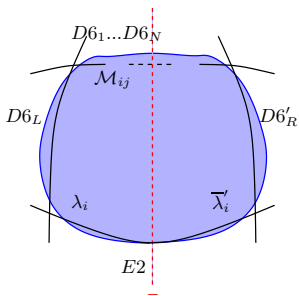}
\caption{Type IIA intersecting $D6$-brane setup involving a $\O$ ``stringy" instanton.}
\label{O1fig1}
\end{center}
\end{figure}

It is clear that the above considerations translate as well to $\O$ D-instantons, except that we do not need anymore a single additional space-time filling brane on the same cycle (which provides a global $U(1)$ symmetry). One can easily write down simple HW setups giving rise to similar dynamics as before, however we refer instead to the a bit more involved examples of \cite{GarciaEtxebarria:2007zv}. Here, we take as a simple and illustrative example an orientifold of the above intersecting $D6$-brane setup with the $D6$-brane on top of the $E2$-brane removed. This is illustrated in figure \ref{O1fig1}.
Clearly, $W_{cl}$ disappears, since we do not have space-time quarks $q_i,\c q_j$ and the meson $\M_{ij}$ is projected to the anti-symmetric 2-tensor representation. The world-sheet disk coupling is projected to $\lambda_i\M_{ij}\lambda_j$ such that the D-instanton will yield a non-perturbative superpotential of the form
\beq\label{O1eq1}
W_{np}\sim Pf(\M).
\eq
The effective low-energy 4D gauge group is then given by
\beq
... \times SU(N_L)\times ``Sp(0)"\times ..., 
\eq
where again we keep the non-existing $``Sp(0)"$ gauge group for illustrative purposes. In this simple example, the non-existing gauge factor together with the non-perturbative superpotential given in equation (\ref{O1eq1}) looks like the magnetic dual of an s-confining electric $Sp(2N_c)$ theory with $N_f=N_c+2$. In fact, in HW examples, one can easily translate such $\O$ D-instanton setups to the dual electric theory, as for example done in \cite{Aharony:2007pr,GarciaEtxebarria:2007zv}. The underlying point is, that again, the $\O$ D-instanton alone can be seen as the instanton of the completely broken $Sp(N)$ gauge group corresponding to the ``empty" cycle wrapped only by the $E2$-brane. Thus, as before we can reproduce in pure field-theory qualitatively the effective superpotential (\ref{O1eq1}) by simply taking the ``empty" cycle with the D-instanton as a completely broken $Sp(2)\sim SU(2)$ gauge group as UV completion into account.

\section{$\SP\times SO(3)$ theory}

Since these considerations imply that at least some class of ``stringy" D-instantons reproduce effects which one could as well construct in pure gauge theory, we should as well be able to obtain the known non-perturbative potential for a $SO(3)$ gauge theory with matter in the vector representation and some singlets \cite{Seiberg:1994pq,Intriligator:1995id}. Note that the instanton effect which generates the non-perturbative potential we are after is in field-theory reminiscent of a broken $SO(4)$ and thus is a prime example for the previous discussions. As is clear, the ``stringy" generation of this potential will envolve a $\SP$ D-instanton, for which so far it was only possible to rederive an ADS like potential with the restriction $N_f=N_c-3$ \cite{Akerblom:2006hx,Bianchi:2007fx,Bianchi:2007wy}. However, it is easy to see that similar as in \cite{Petersson:2007sc}, for $N_c=3$, as in our $SO(3)$ case, the ADHM constraints absorb the excess of uncharged zero-modes since the fermionic part of the integration after implementing the ADHM constraints will be given by
\beq\label{USPeq1}
\int[d\beta^A_c][d\lambda^A_c]\delta^{(3)}(b^A_c\sigma^i_{AB}\beta^B_c)\delta^{(3)}(\c b^A_c\sigma^i_{AB}\beta^B_c)e^{-S_{int}},
\eq
where $\sigma^i$ are the Pauli-matrices, $\beta_c^A,b_c^A,\c b_c^A$ are the fermionic, respectively bosonic,  zero-modes from the strings stretching between the two E2-branes (capital indices) and the three D6-branes (lower-case indices) and $\lambda_c^A$ are the fermionic zero-modes due to strings between the E2-branes and the $N_f$ flavor D6-branes. Expansion of the delta functions yields
\beq
\begin{split}
\delta^{(3)}(b^A_c\sigma^i_{AB}\beta^B_c)\delta^{(3)}(\c b^A_c\sigma^i_{AB}\beta^B_c)=\\
\prod^3_{i,j=1}\sigma^i_{A_iB_i}\sigma^j_{C_jD_j}b^{A_i}_{c_i}\c b^{C_j}_{d_j}\beta^{B_i}_{c_i}\beta^{D_j}_{d_j},
\end{split}
\eq
where the additional subindices indicate that we have to sum over the brane indices separately for each factor.
Thus, we see that the $\beta^A_c$ fermionic zero modes are automatically absorbed and the integral (\ref{USPeq1}) is non-vanishing, if there are suitable couplings which absorb the remaining $\lambda^A_c$ zero modes. In this way, we can reproduce the expected non-perturbative superpotential for the $SO(3)$ under consideration via a ``stringy" D-instanton.

\acknowledgments

It is a pleasure to thank M. Haack, M. Herbst, I. Garcia-Etxebarria, W. Lerche, F. Marchesano and A. M. Uranga for related discussions and as well M. Bianchi and D. L\"ust for comments on the manuscript. This work is supported by an EU Marie Curie EST fellowship.


\end{document}